# Ground State Calculations of Confined Hydrogen Molecule $H_2$ Using Variational Monte Carlo Method

**S. B. Doma[1), F. N. El-Gammal[2)] and A. A. Amer[2)]**

[1)] Mathematics Department, Faculty of Science, Alexandria University, Alexandria, Egypt
E-mail address: sbdoma@yahoo.com
[2)] Mathematics Department, Faculty of Science, Menoufia University, Shebin El-Kom, Egypt

**Abstract**
The variational Monte Carlo method is used to evaluate the ground-state energy of the confined hydrogen molecule, $H_2$. Accordingly, we considered the case of hydrogen molecule confined by a hard prolate spheroidal cavity when the nuclear positions are clamped at the foci (on-focus case). Also, the case of off-focus nuclei in which the two nuclei are not clamped to the foci is studied. This case provides flexibility for the treatment of the molecular properties by selecting an arbitrary size and shape of the confining spheroidal box. An accurate trial wave function depending on many variational parameters is used for this purpose. The obtained results are in good agreement with the most recent results.

## 1. Introduction

The $H_2^+$ molecular ion and $H_2$ molecule are the two simplest molecular systems whose study has rendered important information in the understanding of the electronic and structural properties of larger molecules and constitute the cornerstones of the actual development of molecular physics. During the last years many studies concerning the problem of confined molecular systems have been presented. This is due to the unusual physical and chemical properties observed in such systems when submitted to narrow spatial limitation. When atoms and molecules are confined in either penetrable or impenetrable boundaries their properties undergo significant changes. The topic is of timely interest due to the advent of modern techniques for the synthesis of nanostructured materials such as carbon nanotubes, buckyballs and zeolitic nanochannels which serve as ideal containers for molecular insertion and storage with promising applications [1–4]. Many powerful and sophisticated methodologies (Hartree–Fock, quantum chemical density functional theory, quantum molecular dynamics, to mention a few) have been employed to explore the physical and chemical properties of $H_2$ and other small diatomics within endohedral cages [2, 5–7] and carbon nanotubes [7–10]. For aforementioned systems, various theoretical models have been proposed in the past to analyze confinement effects, on particularly those based on boxed-in molecules. A natural extension of the results presented by Crus *et al*. [11] for confined $H_2^+$ and $HeH^{++}$, Rodriguez *et. al* [12] considered the case of the $H_2$ molecule confined by impenetrable spheroidal boxes when the nuclei do not coincide with the foci. It was shown that by making the cavity size and shape independent of the nuclear positions, optimum equilibrium bond lengths and energies are obtained as compared with corresponding on-focus calculations. This procedure allows for a controlled treatment of molecular properties by selecting an arbitrary size and shape of the confining spheroidal box. One of the most important and recent studies is presented in Ref [13], where a variational approach has been proposed to solve the non-separable Schrödinger problem of the $H_2$ molecule and $H_2^+$ molecular ion in their ground states confined by padded prolate spheroidal cavities. In this study a new treatment was proposed to allow the full control of cavity size and shape, internuclear positions and confining barrier height. The used model adds more flexibility for the treatment of the electronic and vibrational properties of one- and two-electron diatomic molecules when submitted to spatial limitation allowing for a more realistic comparison with experiment.

In this paper a study of the confined hydrogen molecule $H_2$ which is placed inside spherical hard boxes is presented to calculate the ground state energy as function of the characteristics of this hard prolate spheroidal cavity, in the framework of the variational Monte Carlo method, which was used widely to calculate both ground and excited states for unconfined atoms and molecules [14-23].

## 2. Method of the Calculations

Our calculations are based on using variational Monte Carlo (VMC) method which is considered as a one of the most important quantum Monte Carlo methods. It is based on a combination of two ideas namely the variational principle and the Monte Carlo evaluation of integrals using importance sampling based on the Metropolis algorithm [24].

The VMC methods are used to compute quantum expectation values of an operator with a given trial wave function. Given a Hamiltonian operator $H$ and a trial wave function $\psi_T$, the variational principle states that the expectation value is the varitional energy [25]:

$$E_{VMC} = \frac{\int \psi_T^*(\mathbf{R}) \, H \, \psi_T(\mathbf{R}) \, d\mathbf{R}}{\int \psi_T^*(\mathbf{R}) \, \psi_T(\mathbf{R}) \, d\mathbf{R}} \geq E_{exact}, \qquad (2.1)$$

where $\psi_T$ is a trial wave function, $\mathbf{R}$ is the $3N$-dimensional vector of the electron coordinates and $E_{exact}$ is the exact value of the energy of that state. Also, it is important to calculate the standard deviation of the energy [25]

$$\sigma = \sqrt{\frac{\langle E_L^2 \rangle - \langle E_L \rangle^2}{L \, (N-1)}} \qquad (2.2)$$

where $E_L = \frac{\hat{H} \, \psi_T(\mathbf{R})}{\psi_T(\mathbf{R})}$ is the local energy function, $L$ is the ensemble size of random numbers $\{\mathbf{R}_1, \mathbf{R}_2, \ldots, \mathbf{R}_L\}$ and $N$ is the number of ensembles.

## 3. The Hamiltonian of the System

We considered the $H_2$ molecule confined within a prolate spheroidal cavity, defined by the geometric contour $\xi_0$ as shown in Figure-1. The Schrödinger equation for the confined $H_2$ molecule can be written as follows

$$(H - E)\psi = 0, \qquad (3.1)$$

In Figure-1 we presented the geometric characteristics of the confined $H_2$ molecule confined within a prolate spheroidal cavity defined by $\xi_0$. The nuclear charges $Z_a = Z_b = 1$ are both located at distance $d_a$ and $d_b$ from the origin, respectively. $D$ is the interfocal separation and $R$ is the internuclear distance, $R = d_a + d_b$. $r_a$ and $r_b$ are the distances from electron to nuclei $a(b)$, respectively. $r_{1a(b)}$ and $r_{2a(b)}$ are the distances from electrons 1 and 2 to nuclei $a(b)$, respectively, and $r_{1(2)}$, $r^*_{1(2)}$ are their corresponding distances to the foci. $r_{12}$ is the electron–electron distance. From the figure it was easy to write down the Hamiltonian operator corresponding to the coordinates of the two electrons and the two nuclei

$$H = -\frac{1}{2}\nabla_1^2 - \frac{1}{2}\nabla_2^2 - \frac{1}{r_{1a}} - \frac{1}{r_{1b}} - \frac{1}{r_{2a}} - \frac{1}{r_{2b}} + \frac{1}{r_{12}} + \frac{1}{R} + V_c, \qquad (3.2)$$

where $V_c$ is the confining barrier imposed by the spheroidal boundary ($S$), which in our case is assumed to be infinitely high, that is

$$V_c = \begin{cases} \infty & (r_a, r_b) \in S \\ 0 & (r_a, r_b) \notin S \end{cases} \qquad (3.3)$$

In this paper we have considered the case of prolate spheroidal confining box so, we used prolate spheroidal coordinates. It is well known that such a coordinate system consists of families of mutually orthogonal confocal ellipsoids ($\lambda$) and hyperboloids ($\mu$) of revolution. The prolate spheroidal coordinates are defined as [12, 13]

$$\lambda = \frac{r_1 + r_2}{D}, \qquad \mu = \frac{r_1 - r_2}{D}, \qquad \varphi = \varphi \ (\varphi \text{ is the azimuthal angle}). \qquad (3.4)$$

The ranges of these variables are

$$1 \leq \lambda \leq \infty, \qquad -1 \leq \mu \leq 1, \qquad 0 \leq \varphi \leq 2\pi. \qquad (3.5)$$

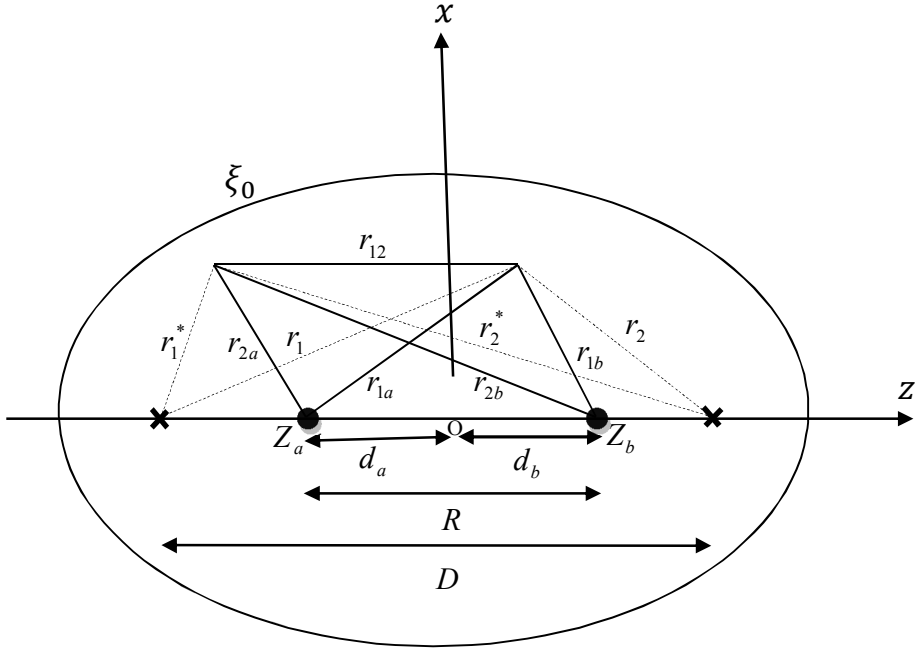

Figure-1 Hydrogen molecule $H_2$ confined within a prolate spheroidal cavity defined by $\xi_0$.

The different sets of coordinates $(\lambda_1, \mu_1, \varphi_1)$ and $(\lambda_2, \mu_2, \varphi_2)$ are assigned, respectively, to electrons characterized by the positions $(r_1, r_2)$ and $(r_1^*, r_2^*)$ relative to the foci as shown in Figure-1. In these coordinates, the kinetic-energy operator and the potential-energy operator are written as [12, 13]

$$-\frac{1}{2}\nabla_i^2 = -\frac{2}{D^2(\lambda_i^2 - \mu_i^2)}\left\{\frac{\partial}{\partial \lambda_i}(\lambda_i^2 - 1)\frac{\partial}{\partial \lambda_i} + \frac{\partial}{\partial \mu_i}(1 - \mu_i^2)\frac{\partial}{\partial \mu_i}\right\} \quad (3.6)$$

and

$$V = -\frac{4}{D}\left(\frac{\lambda_1}{\lambda_1^2 - \mu_1^2} + \frac{\lambda_2}{\lambda_2^2 - \mu_2^2} - \frac{1}{2\rho}\right) + \frac{1}{R} + V_c, \quad (14)$$

respectively, where $\rho = 2r_{12}/D$.

## 4. The Trial Wave Function

In this section we discuss the used trial wave function for solving the Schrödinger equation of free $H_2$ molecule. Our calculations are based on using a trial wave function which was proposed by Kurokawa *et al.* [26]. This trial wave function depends on the Slater type functions which takes the form:

$$\psi = \sum_i C_i (1 + p_{12}) \exp[-\alpha(\lambda_1 + \lambda_2)] \lambda_1^{m_i} \lambda_2^{n_i} \mu_1^{j_i} \mu_2^{k_i} \rho^{l_i}, \quad (4.1)$$

where $p_{12}$ is an electron exchange operator and $C_i$ are the variational parameters which are calculated from the variational principle. This wave function is very simple and similar to the original wave function due to James and Coolidge [27]. James-Coolidge wave function and this wave function differ only in the powers $m_i$ and $n_i$ of the variables $\lambda_1$ and $\lambda_2$: $m_i$ and $n_i$ are always positive in the James-Coolidge wave function, but they can be even negative in this free wave function.

This wave function was constructed from 13 terms; 11 terms of James and Coolidge plus 2 terms within the range of $-1 \leq m, n \leq 1$, $|m - n| \leq 1$, $j \geq 0$, $k \leq 2$, $0 \leq l \leq 1$. The term No. 12 has positive *m* and *n*, but the term No. 13 has negative *m* and *n*. This wave function was used with iterative-complement-interaction (ICI) method to calculate the ground state energy of unconfined hydrogen molecule [26]. In this paper, we employed this type of wave function to study the case of confined hydrogen molecule. For this case, the trial wave function must vanishes at the spherical boundary surface, so a cut-off factor is employed to fulfill this condition and the wave function became

$$\psi = \begin{cases} \sum_i C_i (1 + p_{12}) \exp[-\alpha(\lambda_1 + \lambda_2)] \lambda_1^{m_i} \lambda_2^{n_i} \mu_1^{j_i} \mu_2^{k_i} \rho^{l_i} [(1 - \gamma \lambda_i/\xi_0)] & \text{for } \lambda < \xi_0 \\ 0 & \text{for } \lambda \geq \xi_0 \end{cases} \quad (4.2)$$

In Eq. (4.2) the last factor in parenthesis represents the cut-off factor in terms of the elliptic coordinates and depends on the variational parameter $\gamma$. It guarantees that $\psi(\lambda = \xi_0, \mu) = 0$ at the boundary. This factor has been successfully used in previous variational studies of atoms confined by padded spherical walls [28] and becomes the usual cut-off term for an infinitely hard wall when $\gamma = 1$, as discussed in [12]. This type of cut-off function was found to provide accurate results.

## 5. Results and Discussions

In this paper we have calculated the ground state energies of the $H_2$ molecule using a trial wave function which was introduced in section-4. The variational Monte Carlo method has been employed for the ground state of free and confined $H_2$ molecule. All energies are obtained in atomic units i.e. ($\hbar = e = m_e = 1$) with set of $4 \times 10^7$ Monte Carlo integration points in order to make the statistical error as low as possible. In this section we present the obtained results. On the first hand to gain some confidence on the adequacy of the trial wave function given by Eq. (4.2) for our calculations in frame of VMC method, we first considered the case of unconfined $H_2$ molecule for different values of the internuclear distance $R$. For the free molecule, we clamped the nuclear positions at the nuclei; hence, $D = R$ and $d_a = d_b = R/2$. In Table-1 we compare our obtained results for the behavior of the total energy of the ground state of free $H_2$ molecule with other pervious calculations for a wide set of internuclear distances. For internuclear distances in the range $0.6 \leq R \leq 3.2$, it was sufficient to compare our values with the available SCF in Ref [29], whereas for $4.0 \leq R \leq 8.0$ we compared with results of Ref [30]. Also, the results obtained by Cruz *et al.* [12] are introduced. The results presented in Table-1 indicate clearly that when $R$ increases the electrons interact and become less and less, particularly around the nuclei. Each nucleus has an electron and the probability for both being around the same nucleus is small, as one would expect. When R increases, the $H_2$ molecule tends to separate to two hydrogen atoms in their ground states, therefore the ground state energy decreases. A good quantitative agreement is obtained compared to the corresponding accurate values. These results validate the accuracy of the wave function for calculating the ground state of confined $H_2$ molecule inside a hard prolate spheroidal box.

Table-1 Ground state energy of the free $H_2$ molecule as function of the internuclear distance. In parentheses we show the statistical error in the last figure.

| $R$ | $E_{\text{this work}}$ | $E_{SCF}$ | $E^c$ |
|---|---|---|---|
| 0.2 | 2.2474800(1) | - | 2.2478 |
| 0.4 | -0.0749825(1) | -0.078 693 | -0.0756 |
| 0.6 | -0.7247206(1) | -0.729990[a] | -0.7278 |
| 1.0 | -1.084500(1) | -1.085138[a] | -1.0843 |
| 1.2 | -1.124541(2) | -1.125029[a] | -1.1244 |
| 1.3 | -1.131293(1) | -1.132024[a] | -1.1315 |
| 1.35 | -1.135239(4) | - | -1.1329 |
| 1.375 | -1.133103(1) | -1.133642[a] | -1.133(180) |
| 1.4 | -1.137474(9) | -1.133630[a] | -1.133(181) |
| 1.425 | -1.134544(5) | -1.133379[a] | -1.1329 |
| 1.45 | -1.136994(5) | -1.132908[a] | -1.1325 |
| 1.5 | -1.130748(5) | -1.131375[a] | -1.1310 |
| 1.6 | -1.125689(5) | -1.126352[a] | -1.1259 |
| 2.0 | -1.091085(1) | -1.091648[a] | -1.0911 |
| 2.4 | -1.044736(3) | -1.049331[a] | -1.0488 |
| 3.2 | -0.978582(7) | -0.971512[a] | -0.9704 |
| 4.0 | -0.916097(6) | -0.909130[b] | -0.9102 |
| 6.0 | -0.8208544(7) | -0.819032[b] | -0.8214 |
| 8.0 | -0.7850081(1) | -0.779582[b] | -0.7827 |

[a] Ref [29]. [b] Ref [30]. [c] Ref [12].

Since molecules present different electronic and structural behavior when squeezed into a tiny space in contrast to their free condition; knowledge of the way these changes take place as a function of cavity size, shape and composition is of paramount importance. In this paper, we study the case of hydrogen molecule confined by a hard prolate spheroidal cavity in two cases: when the nuclear positions are clamped at the foci (on-focus) and the case of off-focus nuclei in which the two nuclei are uncoupled from the foci, not clamped at the foci.

Firstly, we study the case in which the nuclei are clamped at the foci $d_a = d_b = D/2$, once a major axis $(D\xi_0, D = R = d_a + d_b)$ is fixed, variation of the internuclear distance $(R = D)$ necessarily implies a change in eccentricity $(1/\xi_0)$, which corresponds to a different cage geometry. In Table-2 we displayed the results obtained for the ground state of the confined $H_2$ molecule within a prolate spheroidal cavity with various major axis $C = R\xi_0$ and different values for the internuclear distance $R$ together with the corresponding accurate variational calculations by Lesar *et al.* [31, 32], Cruz *et al.* [12] and exact QMC calculations by Pang [33]. The agreement with other data is found to be good even for relatively large values of the eccentricity, $(1/\xi_0)$.

Table-2 Ground-state energies for the $H_2$ molecule confined within hard prolate spheroidal boxes with nuclear positions clamped at the foci for selected values of the major axis ($R\xi_0$) as compared with corresponding accurate calculations. In parentheses, we show the statistical error in the last figure.

| $R\xi_0$ | $R$ | $E_{\text{this work}}$ | $E^a$ | $E^b$ | $E^c$ |
|---|---|---|---|---|---|
| ∞ | 1.388 | -1.133296(1) | −1.1332 | - | - |
|  | 1.4010 | -1.173397(1) | - | - | -1.1746 |
| 12 | 1.386 | -1.132001 | −1.1322 | - | - |
|  | 1.403 | -1.156829 | - | −1.1685 | - |
| 10 | 1.372 | -1.128147(4) | −1.1292 | - | - |
|  | 1.395 | -1.162297(2) | - | −1.1638 | - |
| 8 | 1.321 | -1.11102(3) | −1.1102 | - | - |
|  | 1.3503 | -1.1500(3) | - | - | −1.1533 |
| 6 | 1.1771 | -1.0515(2) | - | - | −1.0523 |
|  | 1.208 | -1.040882(1) | - | −1.0441 | - |
| 4 | 0.885 | -0.431810(4) | −0.4321 | - | - |
|  | 0.893 | -0.4744763(6) | - | −0.4749 | - |
|  | 0.8949 | -0.4786076(1) | - | - | −0.4790 |
| 3 | 0.683 | 0.6932845(2) | 0.6934 | - | - |
|  | 0.686 | 0.647240(1) | - | 0.6474 | - |
| 2 | 0.4493 | 4.595042(5) | - | - | 4.5944 |
|  | 0.454 | 4.644851(5) | 4.6433 | - | - |

[a] Ref [12]. [b] Ref [31, 32]. [c] Ref [33].

Secondly, we discuss the case where the two nuclei are allowed to relax out of the focal positions along the major axis. This means that the internuclear distance $R$ and the interfocal distance $D$ have slightly different values from each other. The obtained results for this case are listed in Table-3. Comparing the results obtained in relaxtion case (Table-3) and those of clamped case (Table-2) will reflect the effect of the relaxation of the on-focus nuclei. The comparison ensures that the optimum value of the energy can be obtained when the nuclei do not coincide with the foci. Also, the equilibrium internuclear distances increase relative to the on-focus case with corresponding lowering in the energy. The independence of confining box size and shape on the nuclear positions provide us with additional degree of freedom by controlling the shape and size of the confining box while varying the nuclear positions.

Figure-2 shows in more detail the evolution of the total energy behavior of the ground state energy as a function of $\xi_0$ and the internuclear distance $R$ of $H_2$ molecule enclosed by a prolate spheroidal cavity with major axis $D\xi_0$ for the set of box sizes, considered in this paper, after allowing for nuclear relaxation in the corresponding on-focus calculations.

Table-3 Total energy behavior of the ground state energy of $H_2$ molecule enclosed by a prolate spheroidal cavity with major axis $D\xi_0$ after allowing for nuclear relaxation in the corresponding on-focus calculations. $D$ is the original equilibrium on-focus bond length. In parentheses, we show the statistical error in the last figure.

| $D\xi_0$ | $D$ | $R$ | $E_{\text{this work}}$ | $E^a$ |
|---|---|---|---|---|
| 12 | 1.386 | 1.391 | -1.1328(1) | -1.1322 |
| 10 | 1.372 | 1.376 | -1.1230(1) | -1.1292 |
| 8 | 1.321 | 1.323 | -1.1101(7) | -1.1102 |
| 6 | 1.177 | 1.187 | -1.0069(7) | -1.0081 |
| 4 | 0.885 | 0.913 | -0.4337252(1) | -0.4333 |
| 3 | 0.683 | 0.726 | 0.6875051(1) | 0.6878 |
| 2 | 0.454 | 0.508 | 4.622613(7) | 4.6142 |

$^a$ Ref [12].

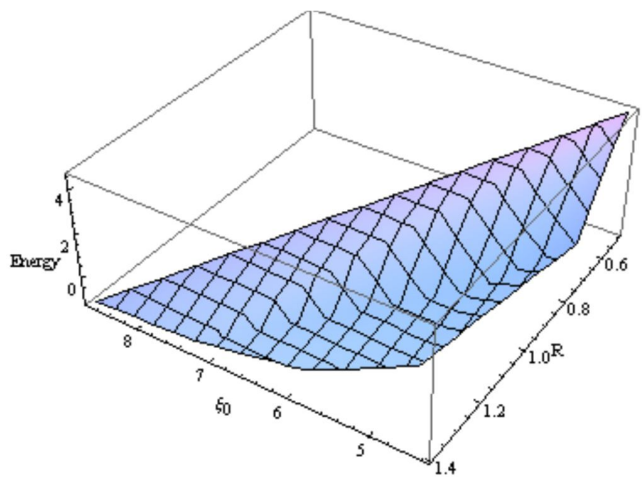

Figure-2 Total energy behavior of the ground state as a function of $\xi_0$ and the internuclear distance $R$ of $H_2$ molecule enclosed by a prolate spheroidal cavity with major axis $D\xi_0$ after allowing for nuclear relaxation in the corresponding on-focus calculations.

On the other hand, we have considered the case of off-focus nuclear positions with fixed eccentricity $e = \frac{1}{\xi_0} = 0.5$. In this case the shape of the cavity will kept fixed where size is variable. The results describing the case in which $H_2$ molecule is compressed within a prolate spheroidal cavity with variable sizes with different values of major axis $C = D\xi_0 = 2, 3, 4, 6, 12$ and different values for the internuclear distance $R$ are displayed in Table-4. In this technique, all boxes keep the same aspect ratio as the volume changes. In Table-4 we compare our results with the first results obtained previously for this case by Cruz *et al*. [12]. It is clear that our results exhibit a good accuracy compared to previous data. The obtained results are presenting for the case of off-focus nuclear relaxation for a fixed confining geometry leading to new energies as compared to the corresponding on-focus calculation.

Table-4 Total energy behavior of the ground state of $H_2$ molecule enclosed by a prolate spheroidal cavity with varying major axis $C = D\xi_0 = 2, 3, 4, 6, 12$ and fixed eccentricity $e = \frac{1}{\xi_0} = 0.5$. In parentheses, we show the statistical error in the last figure.

| $D\xi_0$ | $R$ | $E_{\text{this work}}$ | $E^a$ |
|---|---|---|---|
| 12 | 1.381 | -1.126886(4) | -1.1268 |
| 6 | 1.153 | -0.9311808(2) | -0.9392 |
| 4 | 0.880 | -0.1938857(1) | -0.1938 |
| 3 | 0.704 | 1.212303(2) | 1.2141 |
| 2 | 0.490 | 5.989710(1) | 5.9899 |

[a] Ref [12].

Figure-3 represents the change of the energy versus the internuclear distance $R$ for the free and confined $H_2$ molecule. It is clear that the energy increases in both free and confined cases under decrease of the internuclear distance $R$.

Now, we will introduce a simple chemical analysis concerning the catalytic role of enzyme. Enzymes are macromolecular biological catalysts which play a central role in life due to their catalytic properties. The molecules at the beginning of the process are called substrates and the enzyme converts these into different molecules, called products. The active site is always a non-rigid polar cavity, or crevice, where the substrate will be rearranged in products. There are two cases for converting from the confined state to the free state. Considering a confined molecule with the energy given in point A. In the first case let us assume a sudden release of the constraint, this will relax the bonding electron into the free state with similar internuclear distance $R$. In this hypothesis, the vertical transition of the electrons from A → B (or from confined to unconfined state) leads to change the free molecule state and leave it in a vibrational excited state. In the second case, if the switch-off of the constraint is slower, the relaxation pathway becomes A → C. In this relaxation pathway the nuclei have time to move and so they gain kinetic energy. Hence, the chemical bond is left in a vibrational excited state. In case of strong compression, then molecular bond scission might be obtained however, we can state that in all cases the molecular bond of the free molecule on its electronic ground state is left, at least, in a vibrational excited state. As a result, one can consider the behavior of the vibrational excitation (or bond breaking) to be like the effect of increasing the temperature of the substrate which leads to easier a subsequent atomic rearrangement to give the products. This is a fundamental property, essential for the catalytic role of enzyme.

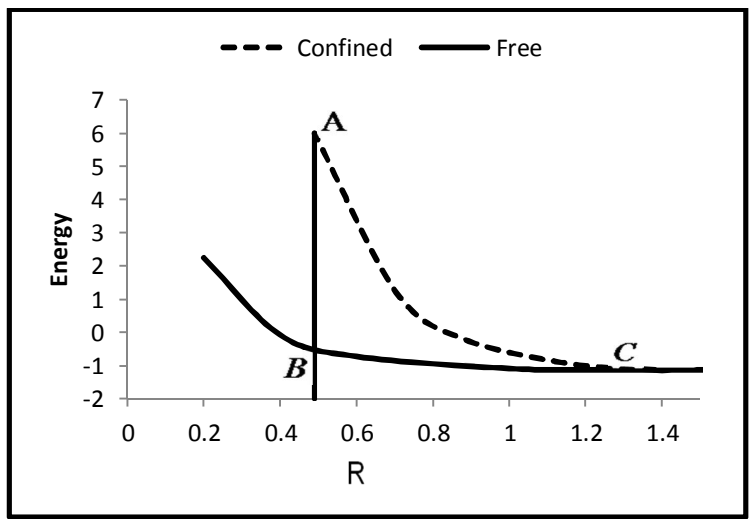

Figure-3 Change of the ground state energy of a confined $H_2$ molecule when the confinement is removed. Two different situations are illustrated.

## 6. Conclusion

In this paper we have used variational Monte Carlo method to study $H_2$ molecule confined within hard prolate spheroidal boxes. We have calculated the energies for both confined $H_2$ molecule and its free case. The ground state energy was plotted as a function of the internuclear distance $R$ of free $H_2$ molecule. It was shown also here that, when the nuclear positions are allowed to relax out of the foci for a fixed cage size and shape, different energies are obtained. Also, the case of off-focus nuclei in which the two nuclei are uncoupled from the foci is studied. In all cases our results exhibit a good accuracy comparing with pervious values obtained using different methods and different forms of trial wave functions. Finally, we conclude that the applications of VMC method can be extended successfully to cover the case of compressed molecules.